# The system of mobile ions in lattice models: Screening effects, thermodynamic and electrophysical properties

George Bokun[a], Dung di Caprio[b], Myroslav Holovko[c], Vyacheslav Vikhrenko[a,*]

[a] Belarusian State Technological University, 13a Sverdlov Str., 220006 Minsk, Belarus
[b] Chimie ParisTech, PSL Research University, CNRS, Institut de Recherche de Chimie Paris (IRCP), 75005 Paris, France
[c] Institute for Condensed Matter Physics, 1, Svientsitskii Street, 79011 Lviv, Ukraine

**Abstract**

The lattice fluid model of the system with short range and long range Coulomb interactions is suggested. In the framework of the collective variables method, the screening of the Coulomb interactions in the bulk is considered. It is shown that the Debye length includes additional concentration dependence inversely proportional to the square root of the mean concentration of vacant sites like what is known at the plane boundary. The Coulomb interaction contribution to the free energy of the system is calculated in the approach close to the mean spherical approximation and is given in an analytical form.

The influence of the variation of the crystal field near the system boundary on the structure and characteristics of the electric double layer is investigated. As compared to the system with equal crystal potentials at the lattice sites throughout the system the pronounced difference for the electric capacitance appears at low absolute values of the surface potential and it is more pronounced for negative electric potentials. The capacitance diverges as the potential values at which the electric field tends to zero and attains negative values in regions of the surface potentials depending on their polarity and values of the surface crystal potential. Negative values of the capacitance may indicate the thermodynamic instability of the system that can result from neglecting the short range interaction contribution.

## 1. Introduction

Solids with high ability of ionic transport are widely used in devices for energy transformation and storage [1–5], electrochemical analysis of compounds, light transformation [6–8], etc. The examples of such solids are superionic conductors, solid electrolytes, intercalation compounds, to name a few. The most prominent feature of such solids is the presence of the subsystem of mobile ions that can move inside of the matrix created by the hosting one, which creates the neutralizing background for moving ions is frequently considered as a continuum that is subject to hydrodynamic equations of motion [9–12], and can be regarded as a specific fluid or liquid. On the other hand, the mobile ions move in the potential field created by the host subsystem that can be imagined as the host subsystem potential relief. The mobile ions perform thermally activated hops between the potential relief minima that in many cases are distributed like sites of a periodic lattice. The hops can be performed to vacant sites. The concept of vacancies first introduced for explaining diffusion in ionic solids and simultaneously extended to liquids [13,14] afterwards was widely used for investigating numerous phenomena in physics of condensed matter including e.g. hole transport in semiconductors.

In the first approximation, it can be assumed that the role of the host is just in the creation of the potential relief, which does not change under the influence of mobile ions movement. Nevertheless, real solid electrolytes have a complex structure. For example, ceramic electrolytes consist of grains and intergrain layers. Numerous models for describing intergrain regions were suggested [15–22]. They include space charge distribution due to impurities segregation at the grain boundaries [15–17], Schottky-type potential barriers at the grain boundaries [18–20], a linear diffusion model [21,22] and others. The Monte Carlo simulation shows an unexpected charge distribution near the grain boundary modeled by potential barriers [23]. Like a grain boundary, the system boundary violates its symmetry and can result in variation of the energy well depths on the potential relief close to the boundary. Below the one component subsystem of mobile ions in the bulk and at a system boundary is considered in the framework of lattice models.

Lattice fluid models are widely used for describing the subsystem of mobile ions in concentrated electrolytes, ionic liquids, solid electrolytes and superionic conductors [24–34]. The main efforts have been concentrated on investigating in the mean field approximation the structure and properties of the electric double layers arising at the influence of an applied external potential. The lattice models permit accounting of spatial restrictions for particles distribution due to their size. These restrictions remind the requirements of the Fermi–Dirac statistics for the occupation numbers. Thus, Fermi-type distributions for ions in the electric double layers were deduced in particular forms, which were model dependent. The less known fact is that the Debye length in the region of the electric double layer contains the difference between the maximal and current occupation numbers in the bulk, as a correction for its concentration dependence [30].

Less attention has been paid to bulk properties of electrolytes especially beyond the mean field approximations although they are important for investigating their electro-physical characteristics and possible phase transitions. The method of collective variables [35–37] is a promising approach for accounting of correlations in the homogeneous bulk as well as in the non-homogeneous electrolytes of different types. On the other hand, the statistical method of conditional distributions [38–40] can help rationalize accounting for the restrictions on spatial distribution of ions. In the framework of these methods, the models of interface boundaries can be constructed for accounting of e.g. specific adsorption of ions.

## 2. Charge screening in the bulk

Let us consider a lattice fluid model of $N$ charged particles (ions) that can move over $M>N$ sites of the lattice created by the host system, which plays a role of neutralizing background as well. The particles of the host system are fixed at their given positions. As examples, we can mention superionic conductors (e.g., AgJ), where cations can move on the background of anions, or Yttrium stabilized Zirconium (YSZ), where the vacancies in the oxygen sublattice can move on the background of the host YSZ. For the moving particles, the host system creates the potential energy relief with minima of $-u_i$ at lattice sites (the subscript $i$ indicates the position of the corresponding lattice site). For a homogeneous system all the minima are equal $u_i = u$.

The distribution of particles over the system volume is described by the distribution functions. The two-particle (binary) distribution function $F_2(\mathbf{q}_i,\mathbf{q}_j)$ can be represented [38–40] through the one-particle (unary) functions $F_1(\mathbf{q}_i)$ and the binary correlation function $g(\mathbf{q}_i,\mathbf{q}_j)$

$$F_2(\mathbf{q}_i,\mathbf{q}_j) = F_1(\mathbf{q}_i)F_1(\mathbf{q}_j)g(\mathbf{q}_i,\mathbf{q}_j), \tag{1}$$

where $\mathbf{q}_i,\mathbf{q}_j$ determine positions of particles near the corresponding lattice sites.

For solid electrolytes, the unary distribution function is a periodic function with sharp extremes at the lattice sites. In the method of collective variables [35] it was shown that in the approximation corresponding to the Debye description of ionic systems, the correlation function can be written through the dimensionless screened Coulomb potential $u(\mathbf{q}_i, \mathbf{q}_j)$:

$$g(\mathbf{q}_i, \mathbf{q}_j) = \exp[-u(\mathbf{q}_i, \mathbf{q}_j)](1+\ldots), \quad \beta = 1/(k_B T), \tag{2}$$

$$u(\mathbf{q}_i, \mathbf{q}_j) \equiv \beta \tilde{u}(\mathbf{q}_i, \mathbf{q}_j) = (r_B / r) \exp(-\kappa r), \quad r = |\mathbf{q}_i - \mathbf{q}_j|, \quad \kappa = 1/r_D, \quad r_B = \beta e^2 / 4\pi\varepsilon\varepsilon_0, \tag{3}$$

where $\tilde{u}$ and $u$ are the Coulomb and dimensionless Coulomb interaction potentials, respectively, $k_B$ the Boltzmann constant, $T$ the absolute temperature, $r_B$ and $r_D$ are the Bjerrum and Debye lengths, correspondingly, $e$ is the charge of particles, $\varepsilon$ and $\varepsilon_0$ are the electric constant and the medium dielectric constant, respectively.

The screening of the Coulomb interaction is determined by the binary distribution function, and in the framework of the method of collective variables, it is shown [35] that the Fourier transformation of the screened potential can be written as

$$u(k) = \frac{1}{M} \frac{(M/V)\nu(k)}{1+(M/V)\nu(k)m_2(k)}, \tag{4}$$

where

$$\nu(k) = r_B / k^2, \quad m_2(k) \equiv m_2(\mathbf{k}, -\mathbf{k}) = \langle \hat{\rho}_\mathbf{k} \hat{\rho}_{-\mathbf{k}} \rangle_0, \tag{5}$$

$\nu(k)$ is the Fourier transform of the Coulomb interaction, $c=N/M$ is the lattice concentration of particles, $V$ is the system volume, $m_2(k)$ is the second cumulant of the density fluctuations, $\langle \ldots \rangle_0$ designates the averaging over the reference distribution. In [35] it was the averaging over the system of hard spheres; in our case it is the averaging over the ideal crystal host system.

The lattice version of the collective variables [41] generalized to take into account particle displacements from the lattice sites reads as

$$\hat{\rho}_\mathbf{k} = \frac{1}{\sqrt{M}} \sum_{i=1}^{M} \hat{n}_i \exp(i\mathbf{k} \cdot \mathbf{q}_i) - \sqrt{M} c \delta_{\mathbf{k},0}, \quad c = \langle \hat{n}_i \rangle_0, \quad \mathbf{q}_i = \mathbf{R}_i + \mathbf{r}_i, \tag{6}$$

where $\hat{n}_i$ is the occupation number operator ($\hat{n}_i = 1$ if the lattice cell $v_i$ is occupied by a particle, $\hat{n}_i = 0$ if the lattice cell is vacant; multiple occupation of a cell is forbidden), $c$ is the mean lattice concentration of particles, $\delta_{\mathbf{k},0}$ is the Kronecker's $\delta$-symbol, $\mathbf{R}_i$ and $\mathbf{r}_i$ are the radius-vectors of the lattice sites and the particle position with respect to its lattice site, correspondingly.

For calculating the second cumulant it is useful to note that

$$\langle \hat{n}_i \exp(i\mathbf{k} \cdot \mathbf{q}_i) \rangle_0 = c \exp(i\mathbf{k} \cdot \mathbf{R}_i) f(\mathbf{k}), \quad f(\mathbf{k}) = \int_{v_i} \exp(i\mathbf{k} \cdot \mathbf{r}_i) F_1(\mathbf{r}_i) d\mathbf{r}_i, \quad \mathbf{r}_i \in v_i, \tag{7}$$

where normalized to 1 unary distribution function is used

$$\int_{v_i} F_1(\mathbf{r}_i) d^3 \mathbf{r}_i = 1, \quad v_i = v = V/M, \tag{8}$$

$v_i = v$ is the volume of a lattice cell.

The Fourier transform of the concentration fluctuation correlation

$$\left\langle \rho_{\mathbf{k}_1}\rho_{\mathbf{k}_2}\right\rangle_0 = \left\langle \frac{1}{M}\sum_{i=1}^{M}\sum_{j=1}^{M}\exp[i(\mathbf{k}_1\cdot\mathbf{q}_i+\mathbf{k}_2\cdot\mathbf{q}_j)]\right\rangle_0 - Mc^2\delta_{\mathbf{k}_1,0}\delta_{\mathbf{k}_2,0}. \tag{9}$$

The facts that $f(0)=1$ and

$$\sum_{i=1}^{M}\exp(i\mathbf{k}\cdot\mathbf{R}_i) = M\delta_{\mathbf{k},0} \tag{10}$$

were used deriving Eq. (9).

The first term in the r.h.s. of Eq. (9) can be split into two sums resulting in

$$\left\langle \frac{1}{M}\sum_{i=1}^{M}\sum_{j=1}^{M}\exp[i(\mathbf{k}_1\cdot\mathbf{q}_i+\mathbf{k}_2\cdot\mathbf{q}_j)]\right\rangle_0 =$$

$$= \left\langle \frac{1}{M}\sum_{i=1}^{M}\sum_{\substack{j=1\\j\neq i}}^{M}\exp[i(\mathbf{k}_1\cdot\mathbf{q}_i+\mathbf{k}_2\cdot\mathbf{q}_j)]\right\rangle_0 + \left\langle \frac{1}{M}\sum_{i=1}^{M}\exp[i(\mathbf{k}_1+\mathbf{k}_2)\cdot\mathbf{q}_i]\right\rangle_0. \tag{11}$$

The first term in the last expression

$$\left\langle \frac{1}{M}\sum_{i=1}^{M}\sum_{\substack{j=1\\j\neq i}}^{M}\exp[i(\mathbf{k}_1\cdot\mathbf{q}_i+\mathbf{k}_2\cdot\mathbf{q}_j)]\right\rangle_0 = \frac{1}{M}c^2 f(\mathbf{k}_1)f(\mathbf{k}_2)\sum_{i=1}^{M}\sum_{\substack{j=1\\j\neq i}}^{M}\exp[i(\mathbf{k}_1\cdot\mathbf{R}_i+\mathbf{k}_2\cdot\mathbf{R}_j)] =$$

$$= c^2 f(\mathbf{k}_1)f(\mathbf{k}_2)\left[M\delta_{\mathbf{k}_1,0}\delta_{\mathbf{k}_2,0} - \delta_{\mathbf{k}_1+\mathbf{k}_2,0}\right]. \tag{12}$$

The first term in the r.h.s. of Eq. (12) cancels with the last term of Eq. (9) and the final result is

$$\left\langle \rho_{\mathbf{k}_1}\rho_{\mathbf{k}_2}\right\rangle_0 = c(1-c)f(\mathbf{k}_1)f(\mathbf{k}_2)\delta_{\mathbf{k}_1+\mathbf{k}_2,0}, \tag{13}$$

and the second cumulant

$$m_2(k) = \begin{cases} c(1-c) & \text{at } k=0, \\ c(1-c)f^2(k) & \text{at } k\neq 0. \end{cases} \tag{14}$$

In the approximation, where the unary distribution function is a spherically symmetric function we find that

$$f(\mathbf{k}) = f(-\mathbf{k}) = f(k) = 1 - Jk^2 + Wk^4 + \ldots \tag{15}$$

where the second and fourth moments of the distribution function are

$$J = \frac{1}{2}\int_{v_l}[(\mathbf{k}/k)\cdot\mathbf{r}]^2 F_1(r)d^3\mathbf{r}, \tag{16}$$

$$W = \frac{1}{4!} \int_{v_l} [(\mathbf{k}/k) \cdot \mathbf{r}]^4 F_1(r) d^3\mathbf{r}. \tag{17}$$

Substituting Eqs. (14) and (15) at $k \neq 0$ into Eq. (4) for the Fourier transform of the screened Coulomb potential one arrives at the expression

$$u(k) = \frac{1}{V} \frac{v(k)}{1 + (N/V)v(k)[(1-c) + c(2Jk^2 - (J^2 + W)k^4) + \ldots]}, \tag{18}$$

that means the Debye length in the $k \to 0$ limit in the lattice version includes in the denominator an additional multiplier $(1-c)$

$$r_D = \sqrt{\frac{\varepsilon \varepsilon_0 h^3}{\beta c(1-c)e^2}}, \qquad \frac{N}{V} = \frac{Mc}{V} = \frac{c}{b^3}, \tag{19}$$

where $b$ is the lattice spacing in the case of a simple cubic lattice.

In accordance with Eq. (19) the Debye screening length is symmetric with respect to $c$ and $(1-c)$ as it should be because the description of solid electrolytes can be performed for real ions of concentration $c$ or the Kröger–Vink notations [28,42] can be used, when lattice defects (vacancies) of concentration $1-c$ are considered. The physical quantity, the Debye screening lengths must not depend on the description used. The Debye screening length tends to infinity in both limits $c \to 0$ and $c \to 1$. The concept of vacancies was used to explain the increase of the screening length in concentrated liquid electrolytes when solvent molecules are considered as specific defects that make inaccessible to ions corresponding regions of the system [43]. This is an example of transfer of solid state representations into understanding the behavior of liquids that were originally introduced in [13,14].

It is worth to note that given in Eq. (19) the nonmonotonic concentration dependence of the screening length is conditioned by the symmetry properties of the ideal entropy contribution in the mean field approximation. This is conceptually different from such an effect in concentrated electrolytes and ionic liquids where it is concerned with creation of cation/anion complexes, or due to accounting of interparticle correlations [44–46].

In a more general case as it follows from Eq. (18) the Fourier transform of the screened Coulomb potential in the denominator includes the wave vector dependent terms that account for the displacement of particles from the lattice sites.

## 3. Thermodynamics of the system

For the configuration integral of the system the identity

$$Q_N = Q_{0N} \left\langle \exp(-\beta U_N^C - \beta(U_N^\phi - U_N^0)) \right\rangle_0, \tag{20}$$

can be used where $U_N^\phi = \sum_{i<j}^N \Phi(|\mathbf{q}_i - \mathbf{q}_j|)$ is the energy of short range interactions, $\beta U_N^C = r_B \sum_{i<j}^N r_{ij}^{-1}$ is the Coulomb interaction energy; $U_N^0 = \sum_{i=1}^M \sum_{j \neq i}^M \varphi_j(\mathbf{q}_{\alpha i})$ and $Q_{0N}$ are the energy and configuration integral of the reference system, respectively, that in our case is taken as an ideal crystal [38,39] and thus

$$Q_{0N} = \left( \prod_{\alpha=0}^{1} Q_\alpha c_\alpha^{-c_\alpha} \right)^M, \quad Q_\alpha = \int_{v_i} \exp[-\beta \sum_{j \neq i}^{M} \varphi_j(\mathbf{q}_{\alpha i})] d\mathbf{q}_{\alpha i}, \tag{21}$$

where $\varphi_j(\mathbf{q}_{\alpha i})$ is the mean potential exerted by a particle in the lattice cell $v_j$ on a particle in the lattice cell $v_i$.

The leading term in the expression for the binary distribution function is calculated in accordance with Eqs. (1) and (2). The averaging over reference distribution results in the equation

$$\left\langle \exp(-\beta U_N^C - \beta(U_N^\phi - U_N^0)) \right\rangle_0 = \left\langle \prod_{i \neq j}^{M} (1 + f_{ij}) \exp(-\beta U_N^C) \right\rangle_0, \tag{22}$$

$$f_{ij} = \exp\{-\beta[\Phi(|\mathbf{q}_j - \mathbf{q}_i|) - \varphi(\mathbf{q}_j) - \varphi(\mathbf{q}_i)]\}, \tag{23}$$

where $f_{ij}$ are the renormalized Mayer functions.

The mean potentials are calculated at the condition that the two-vertex diagrams in the expansion of the configuration integral are equal to zero [47]. This condition leads to results that are equivalent to the quasichemical (or Bethe–Peierls) approximation.

Averaging over the reference distribution at the mentioned condition leads to the expression

$$\left\langle \exp(-\beta U_N^C - \beta(U_N^\phi - U_N^0)) \right\rangle_0 = \left\langle \exp(-\beta U_N^C) \right\rangle_0. \tag{24}$$

Using the notation

$$A_c = -\ln \left\langle \exp(-\beta U_N^C) \right\rangle_0 \tag{25}$$

and Eq. (20) the free energy $A$ of the system per lattice site can be represented by the expression

$$\beta A = \sum_{\alpha=0}^{1} c_\alpha (\ln c_\alpha - \ln Q_\alpha) + A_c, \tag{26}$$

where the sum represents the ideal entropy contribution and the lattice part of the free energy due to the short range interactions. The last term accounts for the correlated part of the Coulomb interaction and in accordance with [35] can be written as

$$A_c = \int_0^\beta \langle u_c \rangle d\beta, \tag{27}$$

where $\langle u_c \rangle$ is the density of the mean Coulomb interaction energy. For a homogeneous bulk, due to local electroneutrality conditions, it is determined by the correlation part of the electric potential

$$\langle u_c \rangle = c^2 \int_{v_1} F_1(\mathbf{q}_1) \sum_{j \neq 1}^{M} \int_{v_j} F_1(\mathbf{q}_j) \frac{r_B}{\beta r} h(\mathbf{q}_1, \mathbf{q}_j) d\mathbf{q}_1 d\mathbf{q}_j. \tag{28}$$

The correlation function can be taken in the form that was used for developing the mean spherical approximation [48,49]

$$h(r) = \begin{cases} -u(r), & r > d \\ -1, & r < d \end{cases} \tag{29}$$

where $d$ is the hard core diameter of the particles.

The sharp maxima of the one-particle distribution functions can be approximated by the Gaussian distribution

$$F_1(\mathbf{q}) = (\alpha/\pi)^{3/2} \exp(-\alpha q^2), \quad \alpha d^2 \gg 1, \quad d < b. \tag{30}$$

Then the correlated part of the mean Coulomb interaction is represented as follows

$$\langle u_c \rangle = c^2 \sum_{j=2}^{M} u_{1j}, \tag{31}$$

$$\langle u_{1j} \rangle = \frac{r_B}{\beta} \left(\frac{\alpha}{\pi}\right)^{3/2} \int_{v_j} \frac{h(r_{1j})}{r_{1j}} \exp(-\alpha q_j^2) dv, \tag{32}$$

where $r_{1j}$ is the distance of $j$-th particle from a chosen lattice site, $q_j$ is the displacement of $j$-th particle from its lattice site.

The integral in Eq. (32) can be calculated in the bipolar coordinate system

$$\langle u_{1j} \rangle = \frac{r_B}{\beta} I_{1j}, \tag{33}$$

$$I_{1j} = \sqrt{\frac{\alpha}{\pi}} \frac{1}{R_{1j}} \int_{R_{1j}-r_0}^{R_{1j}+r_0} h(r) \exp[-\alpha(R_{1j}-r)^2] dr, \tag{34}$$

where $R_{1j}$ is the distance between the first and $j$-th lattice sites, $r_0 = b/2$ is the radius of a lattice cell. At the condition $\alpha d^2 \gg 1$, the shape of the lattice cell has no importance because the main contribution to the integrals in Eqs. (28), (32) and (34) comes from the regions close to the lattice sites.

Taking into account that $h(r)$ is a smooth function of $r$ and is multiplied by the function with a strong maximum at the center of the lattice cell the integral in Eq. (34) can be evaluated by the Laplace's method thus giving

$$I_{1j} = \frac{h(R_{1j})}{R_{1j}}. \tag{35}$$

Passing to a description in terms of a continuous medium and performing spatial integration in accordance with Eq. (29) for $h(r)$ we arrive at the following expression for the correlated part of the Coulomb interaction energy in the mean spherical approximation

$$\langle u_c \rangle = u_{1c} + u_{2c}, \tag{36}$$

where

$$u_{1c} = -\frac{2\pi c^2 r_B d^2}{\beta v}, \tag{37}$$

$$u_{2c} = -\frac{4\pi c^2 r_B^2}{\kappa \beta v} e^{-\kappa d}. \tag{38}$$

Finally, the Coulomb interaction contribution to the free energy follows after integration over temperature

$$A_C = -\frac{2\pi c^2 r_B d^2}{v} - \frac{8\pi c^2 r_B^2}{\kappa^4 v d^3}(2 - e^{-\kappa d}(1+(1+\kappa d)^2)). \tag{39}$$

The concentration dependence of the chemical potential with accounting of the entropy contribution

$$\beta\mu(c) = \ln[c/(1-c)] + \partial A_C / \partial c \tag{40}$$

indicates (Fig. 1) that the homogeneous distribution of mobile charges becomes unstable already at rather low concentration of particles where the concentration derivative of the chemical potential becomes negative. The region of the system stability can be clarified due to more consistent accounting of the short-range interactions and correlations among the mobile ions and especially with the particles of the neutralizing background. The role of charged defect interactions in phase transitions of ionic crystals was discussed in [50,51].

## 4. Charge distribution at a plain boundary

In the first approximation the statistical sum $Q_M$ of a subsystem of mobile (e.g. positive) ions on the lattice of $M$ sites can be calculated from the expression

$$-\ln Q_M = \sum_{i=1}^{M}[c_i \ln c_i + (1-c_i)\ln(1-c_i)] + \frac{r_B}{2}\sum_{i\neq j}^{M}\left(\frac{c_i c_j}{r_{ij}} - \frac{c_i c_j'}{r_{ij}'}\right) + $$

$$+ \frac{\beta}{2}\sum_{i\neq j}^{M}\tilde{h}_{ij}c_i c_j + \frac{\beta}{2}\sum_{i\neq j}^{M}J_{ij}c_i c_j + \frac{\beta}{2}\sum_{i\neq j}^{M}J_{ij}h_{ij}c_i c_j + \beta e\sum_{i=1}^{M}\varphi_i^e c_i - \beta\sum_{i=1}^{M}u_i c_i, \tag{41}$$

where $c_i$ and $c_j'$ are the lattice concentrations of ions and counterions, respectively, $r_{ij}$ and $r_{ij}'$ are the distances between the corresponding ions, $J_{ij}$ the short range interaction potential between ions, $\varphi_i^e$ the external potential, $u_i$ the crystal potential distribution, i.e. the potential energy well depths that are created by the host system for mobile particles.

In Eq. (41) the interparticle correlations are taken into account through lattice version of the binary correlation function $h_{ij}$, which is defined through the expression

$$F_{ij} = c_i c_j g_{ij}, \quad h_{ij} = g_{ij} - 1, \tag{42}$$

and its modification for Coulomb interactions

$$\tilde{h}_{ij} = h_{ij}/r_{ij}. \tag{43}$$

Due to the condition for the weakening of correlations the terms of Eq. (41) containing functions $h_{ij}$ and $\tilde{h}_{ij}$ describe the short range contributions.

The chemical potential follows from differentiation over the lattice concentration

$$\beta\mu_i = -\frac{\partial \ln Q_M}{\partial c_i} = \ln\frac{c_i}{1-c_i} + r_B \sum_{\substack{j=1 \\ j\neq i}}^{M}\left(\frac{c_j}{r_{ij}} - \frac{c'_j}{r'_{ij}}\right) +$$

$$+\beta\sum_{\substack{j=1 \\ j\neq i}}^{M}\tilde{h}_{ij}c_j + \beta\sum_{\substack{j=1 \\ j\neq i}}^{M}J_{ij}c_j + \beta\sum_{\substack{j=1 \\ j\neq i}}^{M}J_{ij}h_{ij}c_j + \beta e\varphi_i^e - \beta u_i. \tag{44}$$

It is assumed here that the concentration dependence of the binary distribution functions $F_{ij}$ is mainly represented by the concentration multipliers in Eq. (42) and the concentration dependence of the correlation functions can be neglected.

In the bulk of the system at zero electric field

$$\beta\mu = \ln\frac{c}{1-c} + r_B \sum_{\substack{j=1 \\ j\neq i}}^{M}\left(\frac{c}{r_{ij}} - \frac{c'_j}{r'_{ij}}\right) + \beta c\sum_{\substack{j=1 \\ j\neq i}}^{M}\tilde{h}_{ij} + \beta c\sum_{\substack{j=1 \\ j\neq i}}^{M}J_{ij} + \beta c\sum_{\substack{j=1 \\ j\neq i}}^{M}h_{ij}J_{ij} - \beta u. \tag{45}$$

The equilibrium condition $\mu_i = \mu$ results in the expression for the deviation $\delta c_i = c_i - c$ of the ion concentration from its bulk value (the concentration of counterions is considered homogeneous through the whole system)

$$\ln\frac{c+\delta c_i}{1-c-\delta c_i}\frac{1-c}{c} + \beta\delta w_i + \beta e\varphi_i - \beta\delta u_i = 0, \tag{46}$$

where the total electric field

$$\varphi_i = r_B \sum_{\substack{j=1 \\ j\neq i}}^{M}\frac{\delta c_j}{r_{ij}} + \varphi_i^e. \tag{47}$$

and the short range interparticle interaction contribution

$$\delta w_i = \sum_{\substack{j=1 \\ j\neq i}}^{M}(g_{ij}J_{ij} + \tilde{h}_{ij})\delta c_j, \quad \delta u_i = u_i - u. \tag{48}$$

The short range interactions account for the interparticle correlations that are similar to that considered in a recent paper [52] where the lattice gas model of liquid electrolytes was used.

The variation of the potential energy well depths $\delta u_i$ can appear near the system boundary and result in special surface states for ions.

The solution of Eq. (46)

$$\frac{\delta c_i}{c} = -\frac{1-\exp[\beta(\delta u_i - e\varphi_i - \delta w_i)]}{1+[c/(1-c)]\exp[\beta(\delta u_i - e\varphi_i - \delta w_i)]} \tag{49}$$

can be rewritten for the concentration distribution

$$c_i = \frac{1}{1+[(1-c)/c]\exp[-\beta(\delta u_i - e\varphi_i - \delta w_i)]}, \qquad (50)$$

that looks like a Fermi–Dirac distribution [24–34] corrected for inclusion of the interparticle interaction contribution and concentration dependent multiplier in front of the exponent. This is a consequence of the occupation number definition to take values 0 or 1.

In fact, Eq. (50) is not an explicit solution for the concentration distribution because it contains the concentration dependence in the r.h.s. through the correlation term $\delta w_i$ representing the difference analog of an integral equation. This is a fundamental difference of Eq. (50) from the conventional Fermi–Dirac distribution derived for independent particles or in the mean field approximation. It can be solved iteratively. The approach suggested in Ref. [52] to solve such type of nonlinear equations through their linearization with subsequent renormalization of the control parameters can be used.

In the limit of small concentration this distribution approaches to the Maxwell–Boltzmann distribution, again corrected for including interparticle correlations

$$c_i = c \exp[\beta(\delta u_i - e\varphi_i - \delta w_i)]. \qquad (51)$$

In the Debye–Hückel approximation

$$\delta u_i = 0, \quad \delta w_i \approx 0, \quad \beta e\varphi_i \ll 1 \qquad (52)$$

Eq. (49) reduces to

$$\delta c_i \approx -c(1-c)\beta e\varphi_i \qquad (53)$$

and using the Poisson equation

$$\Delta\varphi_i = -\frac{e}{\varepsilon\varepsilon_0}\frac{\delta c_i}{h^3} \qquad (54)$$

the second order differential equation for the field distribution is formulated

$$\Delta\varphi_i = \frac{\beta e^2 c(1-c)}{\varepsilon\varepsilon_0 h^3}\varphi_i \qquad (55)$$

that means the Debye length is again given by Eq. (19) and the width of the double layer diverges at low ($c\to 0$) and high ($c\to 1$) ion concentrations.

Similar expression for the Debye length was deduced [30] on the basis of thermodynamic arguments taking into account that the concentration derivative of the chemical potential has to be proportional to the product $c(1-c)$ [more precisely to $c(1-c_{max})$] that was explained by the symmetry of low particle and low lattice defects concentrations.

In the general case the equation for the electric potential

$$\Delta\varphi_i = \frac{c(1-c)e}{\varepsilon\varepsilon_0 h^3}\frac{1-\exp[\beta(\delta u_i - e\varphi_i - \delta w_i)]}{1-c+c\exp[\beta(\delta u_i - e\varphi_i - \delta w_i)]} \qquad (56)$$

can be rewritten for the electric field $E=-d\varphi/dz$, axis $z$ is perpendicular to the system boundary

$$E^2 - E_0^2 = 2\frac{c(1-c)e}{\varepsilon\varepsilon_0 h^3}\int_{\varphi_0}^{\varphi}\frac{1-\exp[\beta(\delta u - e\varphi - \delta w)]}{1-c+c\exp[\beta(\delta u - e\varphi - \delta w)]}d\varphi, \quad (57)$$

where $E_0$ and $\varphi_0$ are the electric field and potential at the system boundary.

The double layer differential capacitance can be evaluated from the expression [30,33]

$$C = \varepsilon\varepsilon_0 \frac{dE_0}{d\varphi_0} = \frac{\varepsilon\varepsilon_0}{2E_0}\frac{dE_0^2}{d\varphi_0}. \quad (58)$$

At the conditions $\varphi=0$ and $E=0$ in the bulk it follows from Eq. (57) that

$$\frac{dE_0^2}{d\varphi_0} = 2\frac{c(1-c)e}{\varepsilon\varepsilon_0 h^3}\frac{1-\exp[\beta(\delta u_0 - e\varphi_0 - \delta w_0)]}{1-c+c\exp[\beta(\delta u_0 - e\varphi_0 - \delta w_0)]}, \quad (59)$$

$$E_0 = \left[2\frac{c(1-c)e}{\varepsilon\varepsilon_0 h^3}\int_0^{\varphi_0}\frac{1-\exp[\beta(\delta u - e\varphi - \delta w)]}{1-c+c\exp[\beta(\delta u - e\varphi - \delta w)]}d\varphi\right]^{1/2}\text{sign}\,\varphi_0. \quad (60)$$

At $\delta u=0$ and $\delta w=0$ Eqs. (50), (60) coincide with those of Ref. [30].

As is mentioned above the well depths $u_i$ can vary in the region close to the system boundary that will result in additional ion redistribution close to the boundary. Positive $\delta u_i$ values correspond to the enrichment of the surface layers by moving ions.

For qualitative estimation of the influence of this surface potential on the double layer characteristics we adopt that $\delta u$ differs from zero in the first surface layer of thickness $b$ only and consider the mean field approximation when the binary distribution function is represented by the product of unary ones and then $\delta w=0$.

The integral in Eq. (60) can be split into two parts from 0 to $\varphi_1$ and from $\varphi_1$ to $\varphi_0$, where $\varphi_1$ is the value of $\varphi$ at $z=b/2$, the position of ion centers in the first surface layer. $\delta u$ contributes in the latter integral and the result of integration can be represented as follows

$$E_0^2 = \frac{2}{(\beta e \tilde{r}_D)^2}\left\{\beta e\varphi_0 + \frac{1}{c}\ln\left(1-c+ce^{-\beta e\varphi_1}\right)+\right.$$

$$\left.+\frac{1}{c}\ln\frac{1-c+Bce^{-\beta e\varphi_0}}{1-c+Bce^{-\beta e\varphi_1}}\right\}, \quad B=\exp(\beta\delta u_0), \quad \tilde{r}_D = \sqrt{\frac{\varepsilon\varepsilon_0 h^3}{\beta c e^2}}, \quad (61)$$

and from Eq. (59) it follows that

$$\frac{dE_0^2}{d\varphi_0} = \frac{2(1-c)}{\beta e \tilde{r}_D^2}\frac{1-B\exp(-\beta e\varphi_0)}{1-c+cB\exp(-\beta e\varphi_0)}. \quad (62)$$

For calculating $\varphi_1$ the first three terms of its expansion in the first surface layer can be used

$$\varphi_1 = \varphi_0 - E_0 b/2 + \Delta\varphi_0 b^2/8, \quad \Delta\varphi_0 = \Delta\varphi\big|_{z=0}, \quad (63)$$

where $\nabla\varphi\big|_{z=0} = -E_0$ is taken into account.

The nonlinear system of Eqs. (61), (63) can be solved by the method of successive approximations taking e.g. $\varphi_1=0.8\varphi_0$ as an initial trail value. Fig. 2 demonstrates that the solution of Eqs. (61), (63) for $E_0$ is a smooth function of the electric potential. Thus, the method of successive approximations provides with the required solution. The differences between the solutions for $B\neq 1$ and $B=1$ are not very pronounced although they increase with increasing the difference of $B$ from one. At $\varphi_0$ close to zero the values of $E_0$ are close to zero as well. In a narrow region of surface potential values close to zero for $B>1$ and $\varphi_0<0$ or $B<1$ and $\varphi_0>0$ the solution for $E_0$ attains imaginary values that means there is no charge distribution in the electrolyte that satisfies Eqs. (49), (61) with the value of $\varphi_1$ found selfconsistently from these equations. The width of this region increases with increasing the difference of $B$ from one.

The differential electric capacity as a function of the surface electric potential at particular values of the parameters for different values of the variation of the crystal surface potential is shown in Fig. 3. The surface crystal potential variation leads to strong deviation of the differential capacitance at low absolute values of the surface potential and for the negative potentials the difference is more pronounced (Fig. 4). The capacitance diverges at values of $\varphi_0$ at which the electric field tends to zero because at $\delta u \neq 0$ the derivative of square of the electric field over the electric potential is not equal to zero. At the same time the derivative of square of the electric field over the electric potential changes its sign at $\beta e\varphi_0=\beta\delta u$ (see Fig. 4; the inset displaces the behavior of this derivative at small values of the surface potential) that results in negative values of the capacitance in the corresponding ranges of the electric potential and can result in thermodynamic instability of the system. This means that there are no solutions for the charge distribution in the system that corresponds to the surface potential values at which the capacitance is negative.

From physical point of view, the negative values of the capacitance and the system instability are due to the competition of the external electric field and internal crystalline field. It should be noted that according to Eq. (48) the short range interparticle interaction contribution $\delta w_i$ depends on the concentration distribution and Eqs. (49) and (50) are nonlinear equations with respect to the concentration that have to be solved selfconsistently. The account of the influence of the charge distribution in the near surface region through the short range interparticle interaction contribution can significantly change the behavior of the system at low absolute values of the surface potential. On the other hand, the possibility of a negative differential capacitance of the double layer and its divergences were discussed in numerous papers (see [53–56] and references therein) and the problem of system stability in the case of the model under consideration requires additional consideration.

## 5. Conclusion

The lattice fluid model of the system with short range and long range Coulomb interactions is suggested. In the framework of the collective variables method, the screening of the Coulomb interactions in the bulk is considered and it is shown that the Debye length includes additional concentration dependence inversely proportional to the square root of the mean concentration of vacant sites. Additionally, the contribution of the thermal ion displacements from the lattice sites to the Fourier transform of the screened Coulomb interaction is estimated. This contribution results in a more complicated wave vector dependence of the screened Coulomb interaction.

The Coulomb interaction contribution to the free energy of the system is calculated in the approach close to the mean spherical approximation and is given in an analytical form. The analysis of the concentration dependence of the chemical potential shows that this contribution can lead to thermodynamic instability of the system already at rather low mobile ion concentrations. The direct account of the interaction with the immobile neutralizing impurities and short range interion correlations can significantly enlarge the stability region of the system.

In the mean field approximation, the mobile ions near the system boundary obey the Fermi-Dirac type distribution. When the interparticle correlations are taken into account, this distribution reminds the difference analog of the integral equation for the concentration distribution. The influence of the variation of the crystal field near the system boundary on the structure and characteristics of the electric double layer is investigated. As compared to the system with equal crystal potentials at the lattice sites throughout the system, the pronounced difference for the electric capacitance appears at low absolute values of the surface potential. For positively charged particles this difference is more pronounced for negative electric potentials.

The capacitance diverges as the potential values at which the electric field tends to zero and attains negative values in the regions of the surface potentials depending on their polarity and values of the surface crystal potential. From mathematical point of view, the capacitance diverges at values of the surface potential at which the electric field tends to zero while the derivative of the square of the electric field over the electric potential is not equal to zero. From physical point of view, the negative values of the capacitance and the system instability are due to the competition of the external electric field and internal crystalline field. Negative values of the capacitance may indicate the thermodynamic instability of the system although more consistent accounting of the short range interaction contribution can considerably change the behavior of the system at low absolute values of the surface potential.

**Acknowledgement**

This project has received funding from the European Union's Horizon 2020 research and innovation programme under the Marie Skodowska-Curie grant agreement No 734276, the Belarusian Republican Foundation for Fundamental Research (grant №Ф16К-061), the State Fund for Fundamental Research of Ukraine (grant №Ф73/26-2017), and the Ministry of Education of Belarus.

Figure captions.

Fig. 1. The chemical potential versus concentration at $\varepsilon=40$, $T=1000$ K, $b=0.4$ nm, $d=0.35$ nm, $r_D=1.1$ nm, $r_B=0.42$ nm.

Fig. 2. The electric field in units $(\beta e \tilde{r}_D)^{-1}$ versus electric potential. $C_0=\varepsilon/r_D$, $B=\exp(\beta\delta u)$, $\varepsilon=40$, $T=1000$ K, $b=0.4$ nm, $c=0.01$, $r_D=1.1$ nm.

Fig. 3. The differential electric capacity versus electric potential for positively charged particles. $C_0=\varepsilon/r_D$, $B=\exp(\beta\delta u)$, $\varepsilon=40$, $T=1000$ K, $b=0.4$ nm, $c=0.01$, $r_D=1.1$ nm.

Fig. 4. The ratio of the differential capacitance at values of $B\neq 1$ to those at $B=1$. $C_0=\varepsilon/r_D$, $B=\exp(\beta\delta u)$, $\varepsilon=40$, $T=1000$ K, $b=0.4$ nm, $c=0.01$, $r_D=1.1$ nm.

Fig. 5. The derivative of square of the electric field over the surface potential. $C_0=\varepsilon/r_D$, $B=\exp(\beta\delta u)$, $\varepsilon=40$, $T=1000$ K, $b=0.4$ nm, $c=0.01$, $r_D=1.1$ nm.

Highlights

The Debye length in the bulk and double layer contains additional concentration dependence
The interparticle correlations intricate the Fermi-Dirac type distribution of ions
The crystal field variation leads to divergences of the differential electric capacity
The negative values of the differential capacitance may result in system unstability

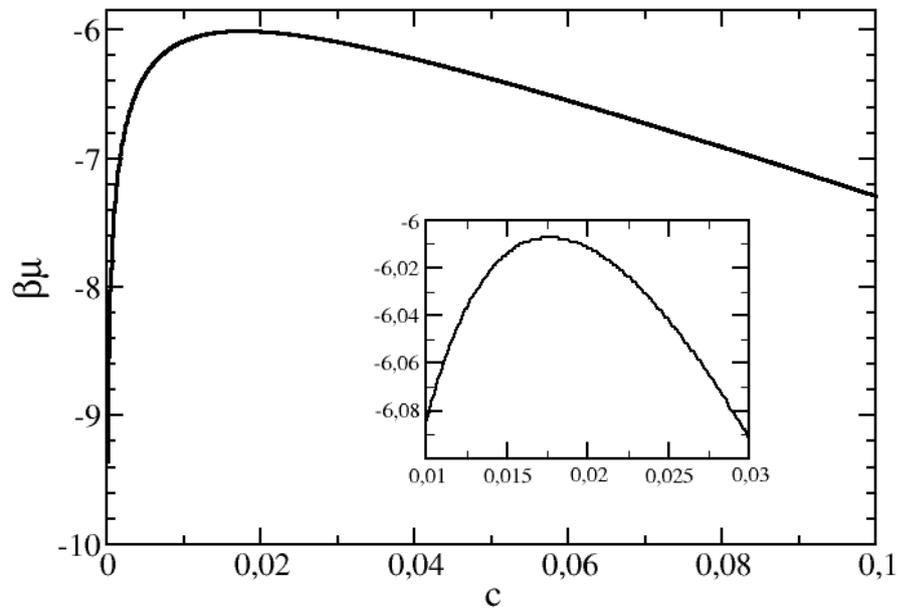

Fig. 1. The chemical potential versus concentration at $\varepsilon=40$, $T=1000$ K, $b=0.4$ nm, $d=0.35$ nm, $r_D=1.1$ nm, $r_B=0.42$ nm.

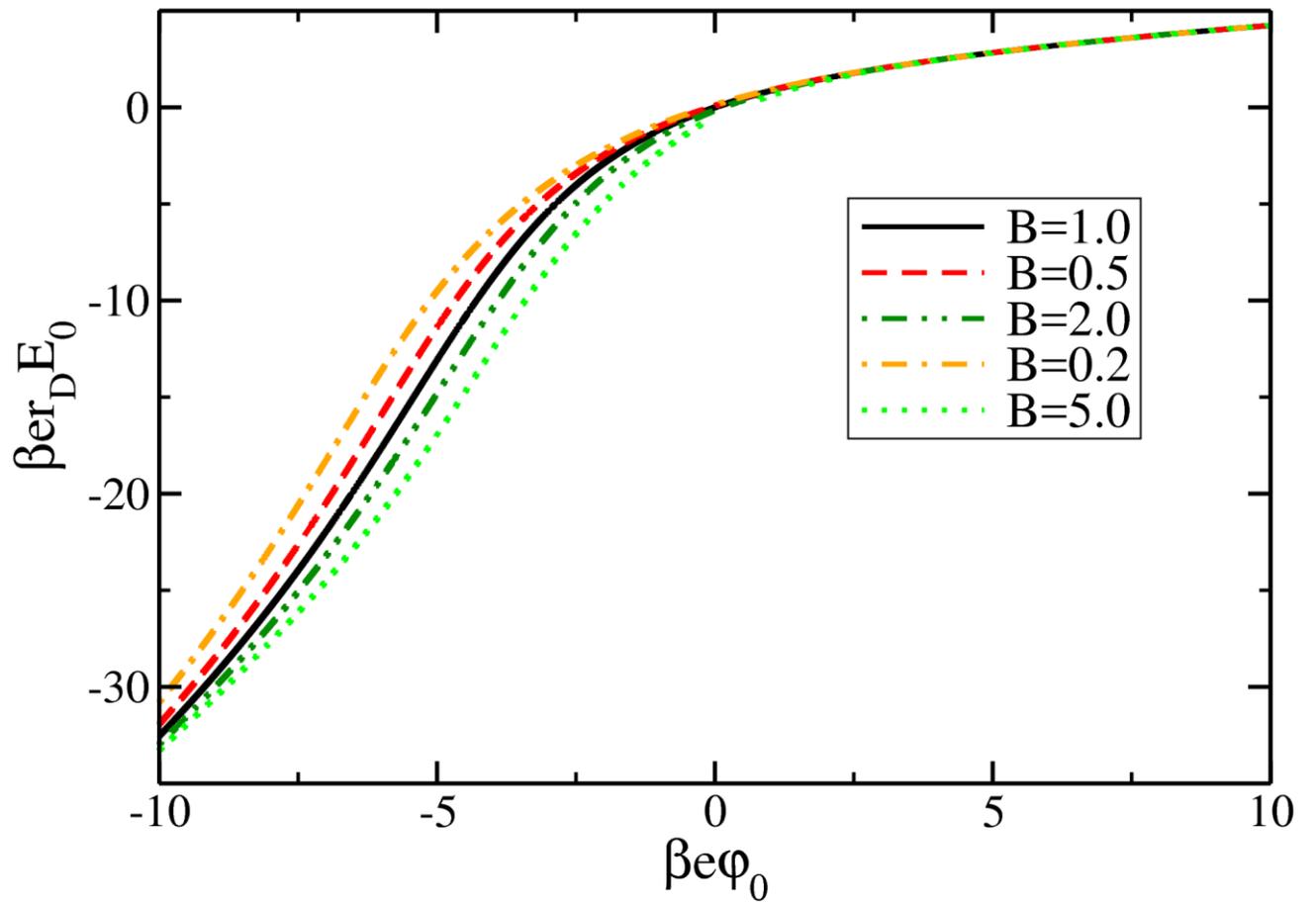

Fig. 2. The electric field in units $(\beta e \tilde{r}_D)^{-1}$ versus electric potential. $C_0=\varepsilon/r_D$, $B=\exp(\beta\delta u)$, $\varepsilon=40$, $T=1000$ K, $b=0.4$ nm, $c=0.01$, $r_D=1.1$ nm.

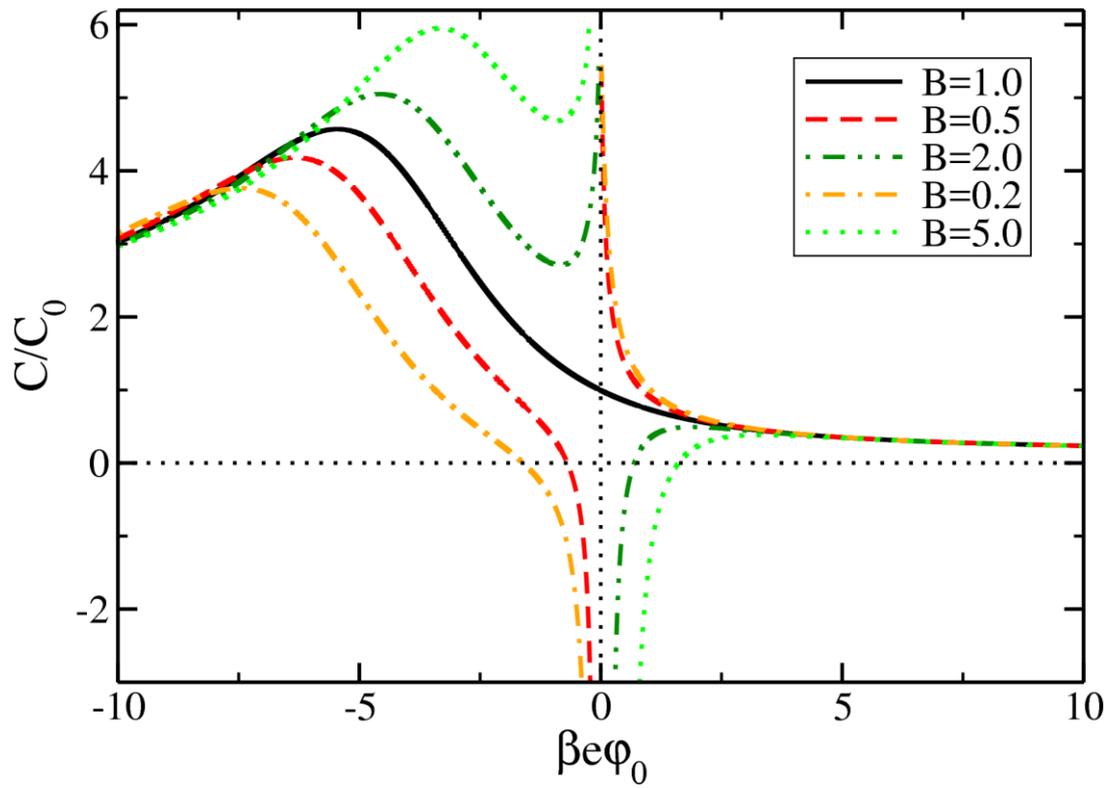

Fig. 3. The differential electric capacity versus electric potential for positively charged particles. $C_0=\varepsilon/r_D$, $B=\exp(\beta\delta u)$, $\varepsilon=40$, $T=1000$ K, $b=0.4$ nm, $c=0.01$, $r_D=1.1$ nm.

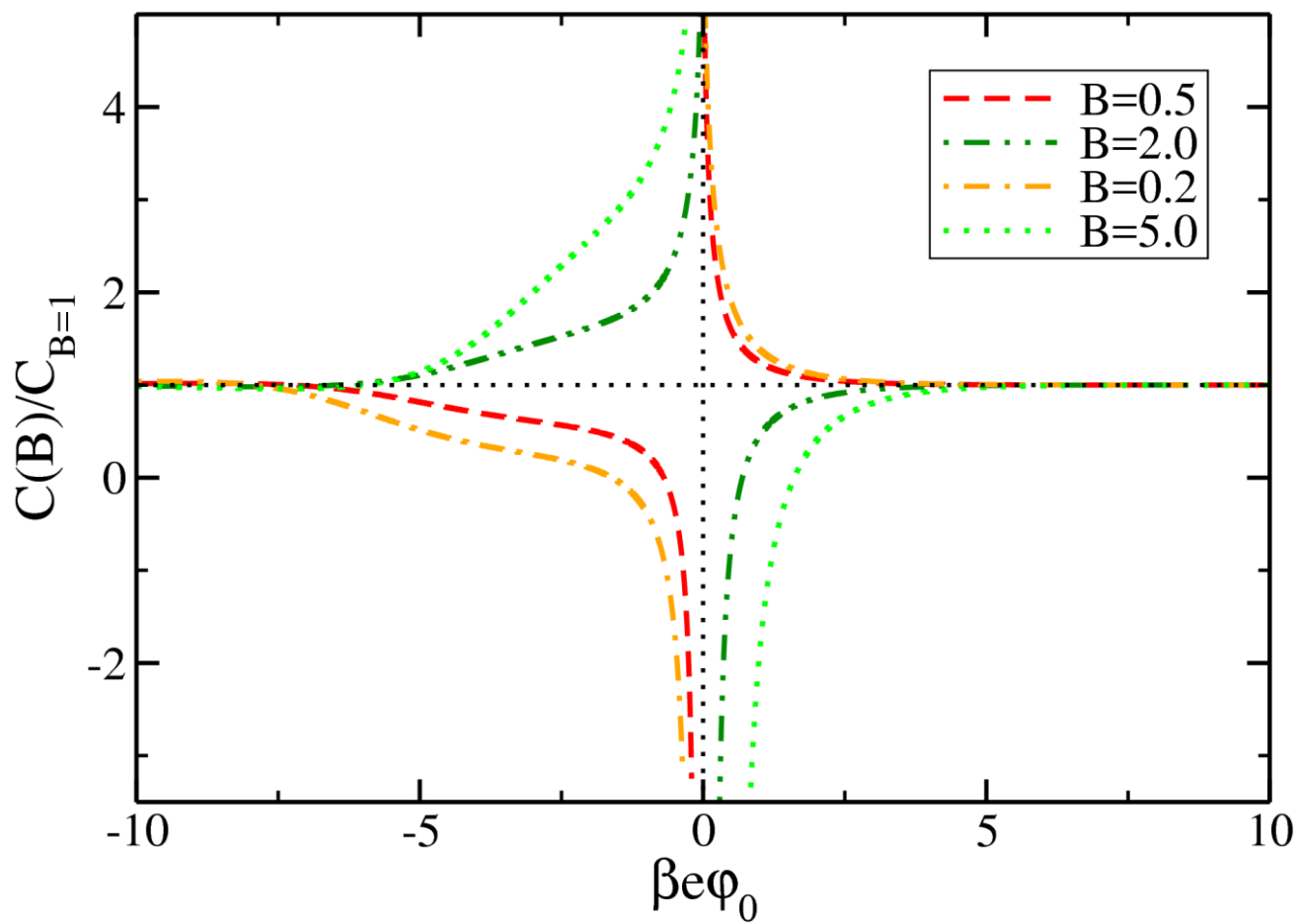

Fig. 4. The ratio of the differential capacitance at values of $B \neq 1$ to those at $B=1$. $C_0 = \varepsilon / r_D$, $B = \exp(\beta \delta u)$, $\varepsilon = 40$, $T = 1000$ K, $b = 0.4$ nm, $c = 0.01$, $r_D = 1.1$ nm.

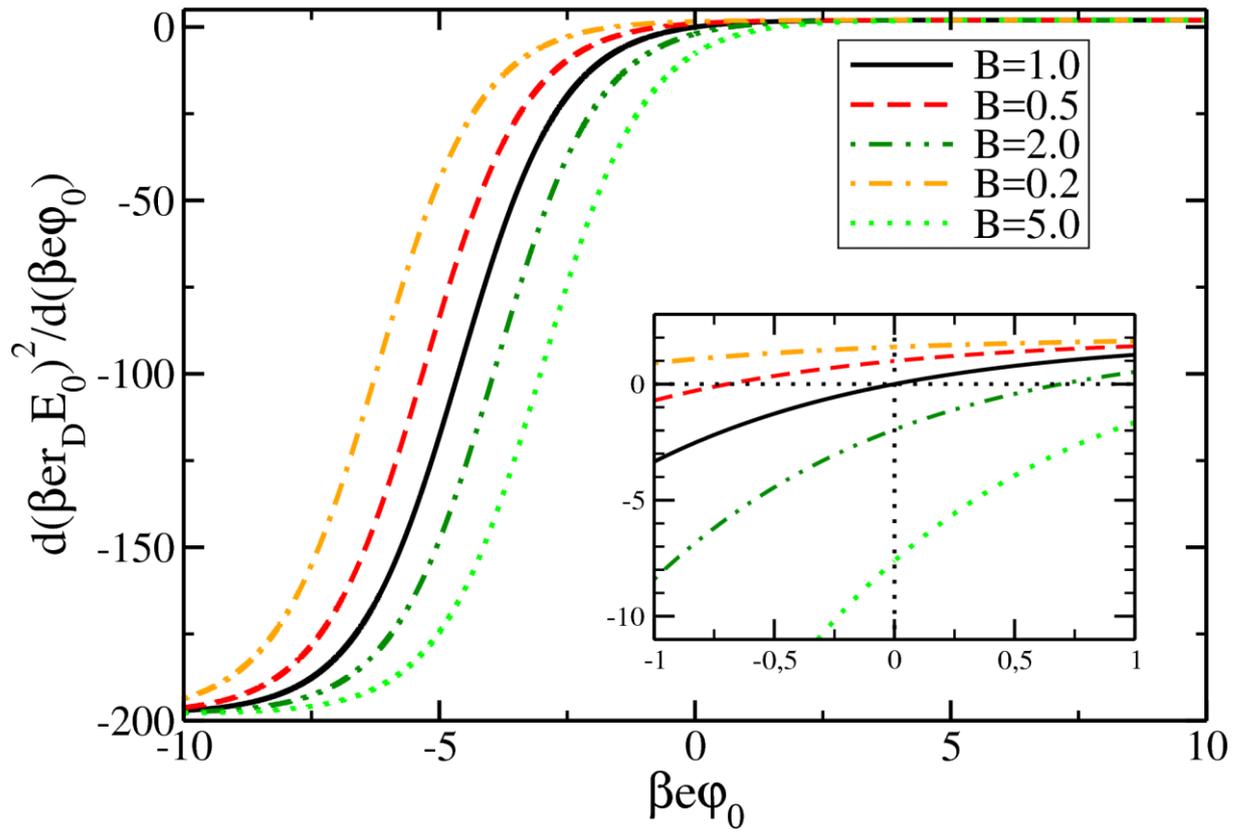

Fig. 5. The derivative of square of the electric field over the surface potential. $C_0=\varepsilon/r_D$, $B=\exp(\beta\delta u)$, $\varepsilon=40$, $T=1000$ K, $b=0.4$ nm, $c=0.01$, $r_D=1.1$ nm.